\documentclass[aps,prl,twocolumn,floats,amsmath,amssymb,showpacs,floatfix]{revtex4}

\usepackage[dvipsnames,rgb,dvips]{xcolor}

\usepackage{graphicx}
\usepackage{dcolumn}

\renewcommand{\cite}{\citep}
\newcommand{\ve}[1]{\ensuremath{\mbox{\boldmath$#1$}}}
\newcommand{\ma}[1]{\ensuremath{\mathbb{#1}}}

\newcommand\nn{\nonumber}
\newcommand{\vvs}{\ensuremath{\ve v_{\rm s}}}
\newcommand{\dl}{\ensuremath{d_{\rm L}}}
\newcommand{\eqnlab}[1]{\label{eq:#1}}
\newcommand{\figlab}[1]{\label{fig:#1}}
\newcommand{\eqnref}[1]{(\ref{eq:#1})}

\newcommand{\Eqnref}[1]{Eq.~(\ref{eq:#1})}
\newcommand{\Figref}[1]{Fig.~\ref{fig:#1}}

\DeclareMathOperator{\E}{{\mathcal F}}
\DeclareMathOperator{\erfc}{erfc}
\DeclareMathOperator{\tr}{Tr}
\DeclareMathOperator{\ku}{Ku}
\DeclareMathOperator{\st}{St}
\DeclareMathOperator{\fr}{{F}}
\DeclareMathOperator{\G}{G}

\begin{document}
\title{Clustering of particles falling in a turbulent flow}
\author{K. Gustavsson, S. Vajedi, and B. Mehlig}
\affiliation{Department of Physics, Gothenburg University, 41296 Gothenburg, Sweden}

\begin{abstract}
Spatial clustering of identical particles falling through a turbulent flow enhances the collision rate between the falling particles, an important problem in aerosol science.  We analyse this problem using perturbation theory in a dimensionless parameter, the so-called Kubo number.  This allows us to derive an analytical theory quantifying the spatial clustering.  We find that clustering of small particles in incompressible random velocity fields may be reduced or enhanced by the effect of gravity (depending on the Stokes number of the particles) and may be strongly anisotropic.
\end{abstract}
% 05.40.-a Fluctuation phenomena, random processes, noise, and Brownian  motion
% 92.60.Mt Particles and aerosols
% 05.60.Cd Classical transport
% 45.50.Tn Collisions
% 47.27.-i Turbulent flows
% 05.40.Jc Brownian motion
% 47.55.Kf Particle-laden flows
% 47.27.eb Turbulence - Statistical theories and models
% 47.27.Gs Isotropic turbulence; homogeneous turbulence
% Find one PACS number of anisotropy, or gravity.

\maketitle

Particles suspended in an incompressible turbulent flow may
cluster together even though direct interactions
between the particles are negligible.
This phenomenon is due to the inertia of the particles.
It has been studied extensively in experiments~\cite{Saw08,Sal08,Gib12}
(see \cite{War09} for a review), in
direct numerical simulations (DNS)~\cite{Chu05,Bec06,Cal08,Men10}, model simulations~\cite{Bec03b}, and by theoretical approaches~\cite{Max87,Dun05,Meh05,Wil07,Gus11a}.
In most DNS, model simulations,  and theoretical studies of clustering,
the effect of gravity is neglected. Those DNS that incorporate gravity
tend to show that clustering is weakened when gravity
causes the particles to fall
through the flow~\cite{Pum04,Wan06,Fra07,Aya08a,Woi09}.
But it has also been reported  that gravity may increase
clustering of particles falling through a turbulent flow~\cite{Woi09},
see also \cite{Pum04}.
\begin{figure}[t]
\includegraphics[width=8cm]{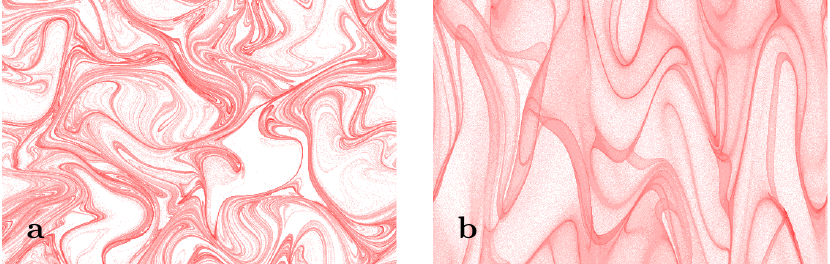} \hspace*{4.cm}
\caption{\figlab{clustering_positions} {\em (Online colour).}
Density of particles falling in a two-dimensional
random velocity field
$\ve u(\ve r,t)=\ve\nabla\psi(\ve r,t)\wedge\hat{\ve e}_3$ in the negative
$y$-direction according to  \Eqnref{eqmStokesGravity}.
Here $\hat{\ve e}_3$ is a unit vector orthogonal to the plane, and
$\psi$ is a Gaussian random function
with zero mean and
$\langle \psi (\ve r,t) \psi (\ve 0,0)\rangle = (u_0^2\eta^2/2)
\exp[-|t|/\tau-\ve r^2/(2\eta^2)]$.
White/red regions show low/high particle densities.
% Particles were initially uniformly distributed with
% velocities equal to the settling velocity $\vvs$, and
% were followed for $100\tau$.
Size of the area shown: $10\eta\times 7\eta$.
Parameters: $\ku=1$, $\fr=1$, $\st=0.2$ ({\bf a}) and $\st=10$ ({\bf b}).
}
\end{figure}

Clustering is commonly explained to be due to the fact
that inertia allows the suspended particles to spiral out from vortices
and gather in straining regions of the flow
(\lq Maxey centrifuge effect'~\cite{Max87}).
This mechanism was derived assuming that
the inertia of the particles is not too large, their
Stokes number  $\st\equiv1/(\gamma\tau)$ must be small.
Here $\gamma$ is the Stokes drag coefficient and $\tau$ is the smallest characteristic time scale of the flow (the Kolmogorov time in turbulent flows).
Even though the fluid-velocity field is incompressible, the 
particle-velocity field acquires a degree of compressibility due to this effect.
The strength of spatial clustering is determined by
the divergence of the particle-velocity field.
Since the explicit dependence upon the gravitational acceleration $\ve g$
drops out when this divergence is taken, it is argued that gravity
does not affect clustering when $\st$ is small.
But gravitational settling affects  the fluctuations of the flow-velocity gradients seen by the falling particles.
In particular, if the gravity parameter \cite{Sha03}
$\fr \equiv |\ve g|\tau/u_0$ is large enough 
($u_0$ is the Kolmogorov speed), 
then gravity pulls particles through the vortices.
How does this affect the spatial clustering of the falling particles?
Do the particles have time to spiral out from the vortices,
or is the Maxey effect destroyed?
What happens at larger $\st$ where \lq preferential sampling\rq{}
in the absence of gravity is strong, but the particle positions
are less correlated with the straining regions of the flow?
How can we explain and quantify the  anisotropy in the spatial patterns
introduced by gravity (\Figref{clustering_positions})?
Finally, the inertial-particle dynamics exhibits
\lq caustics\rq{} \cite{Wil05,Wil06} where the phase-space manifold
that describes the position dependence of
the particle velocities folds over. This gives rise to large
velocity differences between close-by particles
\cite{Fal02,Bec10,Gus11b,Sal12,Bew13,Gus14b}.
How is the rate of caustic formation affected by gravity?
{These open questions are of crucial importance for the process
of rain initiation in warm turbulent rain clouds \cite{Dev12}.}

In order to answer these questions and to quantify the
degree of clustering of particles falling through a turbulent flow
we analyse a model system: particles subject to gravity in a
random velocity field in two spatial dimensions (see \cite{Gus11a}
and caption of~\Figref{clustering_positions}).
We expect no essential difference in three dimensions.
The model has three dimensionless parameters: $\st$, $\fr$, and $\ku$. The Kubo number $\ku\equiv u_0\tau/\eta$ ($\eta$ is the smallest
characteristic length scale of the flow)
is a dimensionless correlation time.
In this Letter we show how to compute the dynamics of the falling particles perturbatively,
taking into account recursively that the perturbations due to the flow velocity
cause the actual particle trajectory to deviate from its deterministic path.
This yields an expansion in $\ku$ \cite{Gus11a,Gus13a,Gus13d},
and results in analytical expressions for the degree of clustering 
and its anisotropy as functions of $\st$, $\fr$, and $\ku$.
Neglecting effects due to finite particle size, we model the dynamics of a
particle as
\begin{align}
\dot{\ve r}=\ku\ve v\,,\hspace{0.5 cm}
\dot{\ve v}=(\ve u(\ve r,t)-\ve v)/\st+\fr\hat{\ve g}\,.
\eqnlab{eqmStokesGravity}
\end{align}
Here dots denote total time derivatives (${\rm d}/{\rm d}t$), $\ve r$ and $\ve v$ are particle
position and velocity,
$\ve u(\ve r,t)$ is the fluid velocity evaluated at the particle position,
and $\hat{\ve g}\equiv\ve g/|\ve g|$ points in the negative $y$-direction.
In \Eqnref{eqmStokesGravity} time- space- and speed scales are de-dimensionalised by the characteristic scales $\tau$, $\eta$, and $u_0$ of the flow.

Preferential sampling
is characterised by the divergence $\ve\nabla\cdot\ve v$ of the particle-velocity field.
We compute the time average of $\ve \nabla \cdot \ve v$  in terms of the matrix
$\ma Z$ of the particle-velocity gradients
$Z_{ij}\equiv\partial v_i/\partial r_j$ \cite{Dun05,Wil07,Gus11a}.
The dynamics of $\ma Z$ follows from \Eqnref{eqmStokesGravity}:
$\dot{\ma Z}=\st^{-1}(\ma A-\ma Z)-\ku \ma Z^2$, and the effect of gravity is implicit
in the dynamics of the matrix $\ma A$ of fluid-velocity gradients
with elements $A_{ij}\equiv\partial u_i/\partial r_j$ evaluated along particle trajectories.
To compute the dynamics of  $\ma A(\ve r_t,t)$ along a particle trajectory $\ve r_t$,
we expand around the deterministic
($\ve u \!=\! \ve 0$)
solution $\ve r_t^{(\rm d)}$ of  \Eqnref{eqmStokesGravity}:
\begin{align}
\ve r_t^{(\rm d)}&=\ve r_0+\ku\vvs t-\ku\st(\ve v_0-\vvs)({\rm e}^{-t/\st}-1)\,.
\eqnlab{rdet}
\end{align}
Here $\ve r_0$ is the initial position of the particle, $\ve v_0$ is its
initial velocity, and $\vvs\equiv \fr\st\hat{\ve g}$ is its settling velocity when $\ve u=0$.
The deviation $\delta\ve r_t\equiv \ve r_t-\ve r_t^{(\rm d)}$
is given by the implicit solution of Eq.~(\ref{eq:eqmStokesGravity}):
 \begin{align}
 \delta \ve r_t &=\frac{\ku}{\st}\int_0^t{\rm d}t_1\int_0^{t_1}{\rm d}t_2e^{(t_2-t_1)/\st}\ve u(\ve r_{t_2},t_2)\,.
 \eqnlab{rsol_implicit}
 \end{align}
For small $\ku$ the deviation $\delta \ve r_t$ is small and
we expand the dynamics of $\ma Z$ and $\ma A$ in powers of
 $\delta \ve r_t$. This results in an expansion of $\tr \ma Z$ in
powers of $\ku$, expressed in terms of products of $\ve u(\ve r_t^{(\rm d)},t)$ and its gradients.
We compute the steady-state average
$\langle\tr\ma Z\rangle_\infty=\langle\ve \nabla \cdot \ve v\rangle_\infty$
using the known statistics of $\ve u(\ve r,t)$.
To order $\ku^3$ we find
\begin{widetext}
\begin{align}
&\langle\ve \nabla \cdot \ve v\rangle_\infty=\frac{3\ku^3}{4\st^5\G^8}\Big\{
2\G^2\st^3(5+4\st+3\st^2-\G^2\st^2(1+\st))
+(1+\st)^3(2(1+\st)^2-\G^2\st^2(\st-3))\E\left[\frac{1+\st}{\sqrt{2}\st\G}\right]^2
\nn\\&
-\sqrt{2}\G\st^2(13\!+\!17\st\!+\!15\st^2\!\!+3\st^3\!\!+\!\G^2\st^2(4\!-\!\st\!-\!3\st^2)\!+\!\G^4\st^4)\E\Big[\frac{1\!+\!\st}{\sqrt{2}\st\G}\Big]
\!-\!4\G\st(1\!+\!\st^2(2\!+\!\st^2\!+\!\G^2))\E\Big[\frac{1}{\G}\Big]
\nn\\&
-2\sqrt{\pi}(1+\st^2)\G(-2+\st^2(-2+(-3+\st^2)\G^2))\int_0^\infty{\rm d}t\exp
\Big[{\G^{-2}}-{t}/{\st}-{\G^2t^2}/{4}\Big]\erfc\Big[\G^{-1}+{\G t}/{2}\Big]
\Big\}\,,
\eqnlab{lambdaSum_Gravity}
\end{align}
\end{widetext}
where $\E[x]\equiv \sqrt{\pi}e^{x^2}\erfc(x)$
and $\G\equiv\ku\st\fr$ ($\ku$ is small but $\G$ can take any value).
Details are given in the supplemental material~\cite{supp}.%, see also \cite{Gus11a,Gus13a,Gus13d}.
% We now discuss the implications of \Eqnref{lambdaSum_Gravity}.

{\em Preferential concentration.}
A series expansion of \Eqnref{lambdaSum_Gravity} to lowest order in $\st$ with $\G$ treated as an independent parameter gives~\cite{supp}:
\begin{align}
\langle\ve \nabla \cdot \ve v\rangle_\infty &
\sim{3\ku^3\st^2}/({4\G^5})
\nn\\&
\times(4\G-6\G^3-(4-4\G^2+3\G^4)\E[\G^{-1}])\,.
\eqnlab{lambdaSum_smallSt}
\end{align}
\Eqnref{lambdaSum_smallSt} describes preferential sampling due to the Maxey centrifuge effect. In the limit $\G\to 0$ \Eqnref{lambdaSum_smallSt} 
approaches
$\langle\ve \nabla \cdot \ve v\rangle_\infty \sim-6\ku^3\st^2$,
see Ref.~\cite{Gus11a}.
Expanding \Eqnref{lambdaSum_Gravity} for small values of $\G$ 
but arbitrary values of $\st$ yields:
\begin{align}
\eqnlab{lambdaSum_smallz}
&\langle\ve \nabla \cdot \ve v\rangle_\infty
 = -6\ku^3\st^2\frac{1+3\st+\st^2}{(1+\st)^3}
\\
&+9\ku^3\G^2\st^2\frac{1+5\st+12\st^2+20\st^3+4\st^4}{(1 + \st)^5}
+\dots\,.
\nn
\end{align}
\Eqnref{lambdaSum_smallz} shows that gravitational settling
reduces preferential sampling for all values of $\st$ provided $\G$ and $\ku$ are small.
We attribute this reduction to the fact that gravity causes the
particles to fall through structures in the flow, rendering their
preferential sampling less efficient.
% We note, however, that sinks in compressible flows
% may act as particle traps for falling particles,
% making preferential sampling more efficient when gravity is present.
% We have confirmed this observation by generalising
% \Eqnref{lambdaSum_Gravity} to compressible flows.
Finally, for large $\G$ and $\st$ \Eqnref{lambdaSum_Gravity} approaches
\begin{equation}
\label{eq:m2}
\langle\ve \nabla \cdot \ve v\rangle_\infty\sim-3\sqrt{2\pi}/\ku^3\st/(4\G^3)\,.
\end{equation}
The full analytical result (\ref{eq:lambdaSum_Gravity}) is compared to results of numerical
simulations of \Eqnref{eqmStokesGravity} in \Figref{lambda_sum}{\bf a}.
We observe good agreement except when $\G$ is small and $\st$ is large.
\begin{figure}[t]
\includegraphics[width=8.5cm]{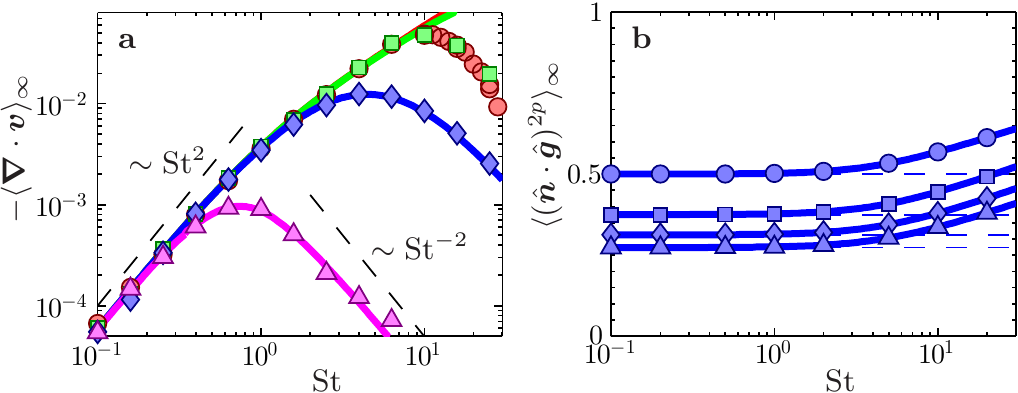}
\caption{\figlab{lambda_sum} {\em (Online colour).}
Average divergence of particle velocity field $\langle\ve \nabla \cdot \ve v\rangle_\infty$, and moments of the alignment
$\langle(\hat{\ve R}\cdot\hat{\ve g})^{2p}\rangle$
between $\hat{\ve R}$ and $\hat{\ve g}$
shown as functions of $\st$ for $\ku=0.1$.
Markers show data from numerical simulations of \Eqnref{eqmStokesGravity}.
{\bf a}
Solid lines show \Eqnref{lambdaSum_Gravity}.
Dashed lines show the slopes of the asymptotes \eqnref{lambdaSum_smallSt} with $\G=0$, and (\ref{eq:m2}).
Parameters: $\fr=0$ (red,$\circ$), $\fr=0.1$ (green,$\Box$), $\fr=1$ (blue,$\Diamond$), $\fr=10$ (magenta,$\triangle$).
{{\cal b} Solid lines show results from Pad\'e-Borel resummation of \eqnref{moments_ndotg} extended to order $\G^{78}$, dashed lines show the
isotropic result. Parameters: $\fr=1$, $p=1$ ($\circ$), $p=2$ ($\Box$), $p=3$ ($\Diamond$), $p=4$~($\triangle$).}
}
\end{figure}

\begin{figure}
\vspace{-0.11cm}
\includegraphics[width=8.5cm]{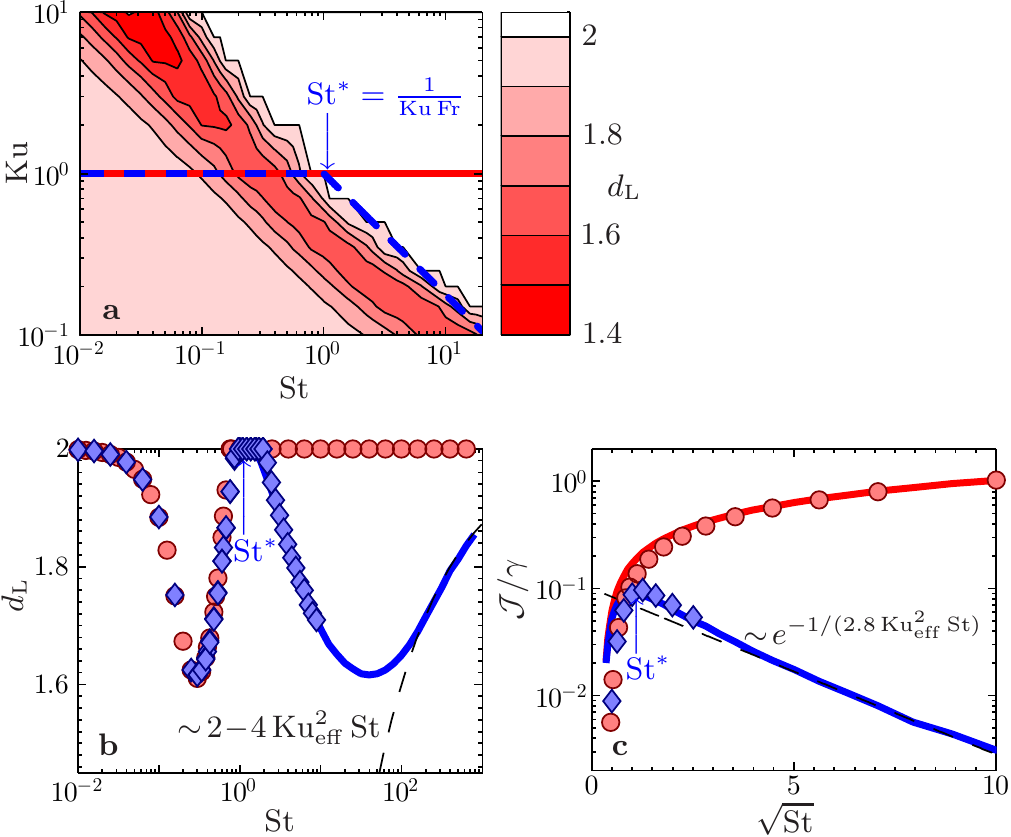}
\caption{\figlab{clustering_gravity} {\em (Online colour).}
Fractal dimension $\dl$ and caustic formation rate $\cal J$ 
from numerical simulations of \Eqnref{eqmStokesGravity}.
{\bf a} Contour plot of $\dl$ as a function $\st$ and $\ku$ for $\fr=0$.
Lines show $\ku_{\rm eff}$, \eqnref{KuEff}, as a function of $\st$ for $\fr=0$ (solid red) and for $\fr=1$ (dashed blue).
{\bf b} shows $\dl$ and {\bf c} shows $\cal J$
versus $\st$ with $\ku=1$ for $\fr=0$ (red,$\circ$) and $\fr=1$ ({blue,$\Diamond$}).
Black dashed lines show limiting behaviors for large $\G$ in terms of the parameter $\ku_{\rm eff}^2\st$ with fitted prefactors.
Solid lines show results of numerical integration of Eqs.~\eqnref{langevin} and \eqnref{diffusion_matrix}.
}
\end{figure}

{\em Preferential alignment.}
\Figref{clustering_positions}{\bf b} shows vertically extended structures compared to the more isotropic structures in \Figref{clustering_positions}{\bf a}.
We now show that this can be explained by the fact 
that the separation vector $\ve R$ between two nearby particles tends to align with $\pm \hat{\ve g}$ for large values of $\G$.
We characterise the alignment between $\ve R$ and $\hat{\ve g}$ by the moments $\langle(\hat{\ve R}\cdot\hat{\ve g})^{2p}\rangle_\infty$ for $p=0,1,\dots$ and $\hat{\ve R}\equiv\ve R/|\ve R|$. Odd-order moments vanish.
% Because two falling particles are interchangeable (the dynamics of $\hat{\ve R}$ 
% is invariant under a sign change), all odd-order moments must vanish.
To find the even moments we use the small-$\ku$ expansion described above. 
The expansion for $\hat{\ve R}$ contains secular terms as well as terms depending
on the initial configuration through $(\hat{\ve R}_0\cdot\hat{\ve g})^p$, 
$\ve v_0$, $\ma Z_0$, $\ve u_0$, and so forth.  Requiring these terms to vanish 
(see~\cite{supp}) yields a recursion which we solve as a power series in $\G$. 
To fourth order in $\G$:
\begin{align}
\!\!\!\langle(\hat{\ve R}\cdot\hat{\ve g})^{2p}\rangle_\infty&\!=\!
\frac{(2p\!-\!1)!!}{2^pp!}\left[1\!+\!\frac{p\G^2}{p\!+\!1}-\frac{p(41\!+\!19p)\G^4}{4(p\!+\!1)(p\!+\!2)}\right]\!.\!
\eqnlab{moments_ndotg}
\end{align}
When $\G=0$ the moments in \Eqnref{moments_ndotg} equal those of isotropically distributed orientation vectors $\hat{\ve R}$. 
For large values of $\G$, $\langle(\hat{\ve R}\cdot\hat{\ve g})^{2p}\rangle_\infty$ increases: $\hat{\ve R}$ aligns with $\pm \hat{\ve g}$ {because the matrix  $\ma Z$
that changes the orientation of $\hat{\ve R}$  approximately follows
an Ornstein-Uhlenbeck process with anisotropic driving.}
As $\G\to\infty$ $\langle(\hat{\ve R}\cdot\hat{\ve g})^{2p}\rangle_\infty\rightarrow 1$ corresponding to complete alignment. 
We have obtained the expansion \Eqnref{moments_ndotg}
to order $78$ in $G$. Pad\'e{}-Borel resummation of this series yields
results in excellent agreement with results from numerical simulations,
\Figref{lambda_sum}{\bf b}. 

{\em Clustering of rapidly falling particles.}
Particles with large settling speeds  are insensitive to instantaneous fluid configurations.
One might expect them to fall uniformly distributed.
But \Figref{clustering_positions}{\bf b} shows that
rapidly falling particles may cluster 
despite the fact that they fall too fast to preferentially sample the flow.
We quantify the degree of clustering by the spatial Lyapunov exponents
\begin{align}
\lambda_1\equiv\lim_{t\to\infty}t^{-1}\ln R_t\,,\quad
\lambda_1+\lambda_2\equiv\lim_{t\to\infty}t^{-1}\ln {\cal A}_t\,,
\eqnlab{lambda_sum_def}
\end{align}
the expansion (contraction) rates of
the spatial distance $R_t\equiv|\ve R_t|$ between two initially nearby
particles, and of the area element ${\cal A}_t$ spanned by the separation
vectors between three nearby particles
\cite{Som93,Bec03b,Wil03,Meh04,Dun05,Wil07}. The Lyapunov exponents
characterise the spatial distribution of the particles; they
form a fractal with dimension~\cite{Kap79}
\begin{align}
\dl\equiv 2-(\lambda_1+\lambda_2)/\lambda_2
\eqnlab{d1_def}
\end{align}
(assuming  $\lambda_1>0$ and $\lambda_1+\lambda_2<0$).
We now show how to evaluate \Eqnref{d1_def} in the limit of
rapidly settling particles (\Figref{clustering_positions}{\bf b}),
assuming that $\G \gg 1$. The deterministic settling trajectory \eqnref{rdet} for large values of $t$
is $\ve r_t^{({\rm d})}=\G t$, and the particles experience rapid fluctuations of the flow velocity
$\ve u(\ve  r_t^{({\rm d})},t)$. {The fluctuations decorrelate rapidly because particles fall many correlation
lengths in a correlation time \cite{Bez06,Fou08},} and we find that the increments of the separation
$\ve R$ and the relative velocity $\ve V$ between two nearby particles follow a
Langevin equation:
\begin{align}
\delta   \ve R = {\ve V}'\, \delta  t'\,,\hspace{0.5cm}
\delta   {\ve V}' = -{\ve V}'\, \delta  t' + \delta \ve F\,.
\eqnlab{langevin}
\end{align}
Here we rescaled $t=t'\st$ and ${\ve v}=\ve v'/(\ku\st)$
as convenient in the
white-noise limit \cite{Dun05,Meh05,Wil07,Gus13a}.
Further  $\delta  \ve F$ is Gaussian white noise with zero mean and
variance $\langle \delta F_i \delta F_j\rangle
= 2 \delta t' \ku^2\st \sum_{kl}D_{ik,jl}
R_{k} R_{l}$. The non-zero $D_{ik,jl}$ are given by
\begin{align}
\nn
D_{21,21}&=\frac{3}{\sqrt{8}G}\E\left[\frac{1}{\sqrt{2}\G}\right]\,,\,
D_{12,12}=\frac{\G^2-1}{2\G^{4}}+\frac{D_{21,21}}{3\G^4}\,,\\
D_{11,11}&=D_{22,22}=-D_{11,22}=-D_{22,11}=-D_{12,21}
\nn\\&
=-D_{21,12}=\frac{1}{2\G^{2}}-\frac{D_{21,21}}{3\G^2}\,.
\eqnlab{diffusion_matrix}
\end{align}
Since $D_{12,12}\ne D_{21,21}$ we see that gravity breaks isotropy, as discussed above.
In \Figref{clustering_gravity}{\bf b} we show $\dl$ obtained by
numerical integration of Eqs.~\eqnref{langevin} and~\eqnref{diffusion_matrix}.
We observe strong clustering at large Stokes numbers, in good agreement with the results
of direct numerical simulations of Eq.~(\ref{eq:eqmStokesGravity}). For $\fr=0$ inertial
particles do not cluster at large $\st$.

A qualitative explanation of this surprising phenomenon goes as follows.
\Figref{clustering_gravity}{\bf a} shows the degree of clustering as a function of $\st$ and $\ku$ for $\fr=0$.  
In this case, and for large values of $\st$, the dynamics of $\ma Z$ is given 
by a single dimensionless parameter, $\epsilon^2\equiv\ku^2\st/2$~\cite{Wil07}.
Likewise, for the case of $\fr\ne 0$ with $\G\gg 1$, the $\ma Z$-dynamics 
turns out to be governed by a single dimensionless parameter, obtained by a change of coordinates that diagonalises
the noise in \Eqnref{diffusion_matrix}. This parameter is
$\epsilon^2/\G^{3/2}=\sqrt{\ku}/(2\sqrt{\st}\fr^{3/2})$, c.f. the parameter dependence of \Eqnref{m2}: 
$\langle\ve \nabla \cdot \ve v'\rangle_\infty\sim-3\sqrt{2\pi}(\epsilon^2/\G^{3/2})^2$
[here $\ve v'$ is the particle velocity in the units used in Eq.~(\ref{eq:langevin})].
For a given value of $\st$ we map the $\fr>0$ dynamics at large $\G$ onto the $\fr\!=\!0$-dynamics by defining an 
effective Kubo number so that   $\ku_{\rm eff}^2\st/2=\sqrt{\ku}/(2\sqrt{\st}\fr^{3/2})$:
\begin{align}
\ku_{\rm eff}\sim
\left\{
\begin{array}{ll}
\ku  & \mbox{if $\st\ll\st^*$}\cr
{\ku^{1/4}/(\fr\st)^{3/4}} & \mbox{if $\st\gg\st^*$}
\end{array}
\right.
\,.
\eqnlab{KuEff}
\end{align}
Here {$\st^*=1/(\ku\fr)$} is the scale at which the large-$\G$ asymptote meets the $\G=0$ asymptote.

Following the curve $\ku_{\rm eff}$ for $\ku=1$ shown in \Figref{clustering_gravity}{\bf a},
the clustering is approximately unmodified for $\st<\st^*$ [see \Figref{clustering_gravity}{\bf b}].
When $\st>\st^*$ the effective Kubo number rapidly becomes so small that the curve reenters the region in parameter space where clustering occurs.
In this limit the clustering is caused by many independent random accelerations
(\lq multiplicative amplification') \cite{Gus11a}, the instantaneous
fluid configuration plays no role.
% The example $\ku=\fr=1$ shown in \Figref{clustering_gravity}{\bf a} generalises to other parameter values. \Eqnref{KuEff} shows that
% larger values of $\ku\fr$ alter $\st^*$ to shift the curve $\ku_{\rm eff}$ towards smaller values of $\st$.
% Larger values of $\ku$ shift the curve towards larger $\ku_{\rm eff}$.
% In the white-noise limit, clustering is strongest near
% $\ku\sim 0.3\st^{-1/2}$ (Fig.~1 in \cite{Wil10b}). This
% line is crossed for large enough values of $\st$ if $\st^*>0.1\ku^{-2}$, i.e. if $\fr<10\ku$.
For large $\fr$, the curve turns early and clustering may be small but non-zero for all values of $\st$ (c.f. $\fr=1$  and $\fr=10$ in \Figref{lambda_sum}{\bf a}).

We remark that due to the anisotropy introduced by $\hat{ \ve g}$, the matrix structure
of the diffusion matrix \Eqnref{diffusion_matrix}  differs from that of the $\fr=0$-case.
This makes the mapping to the effective $\ku$-number approximate.
% When $\ve G<1$ then the situation is more complicated
% because the correlation functions sampled by the falling particles
% are modified. This is reflected in \Eqnref{lambdaSum_Gravity}.

{\em Caustics.} When caustics are frequent, the expansion leading to \Eqnref{lambdaSum_Gravity} and related
expansions in the white-noise limit do not converge, as seen in \Figref{lambda_sum}{\bf a} for large values of $\st$ and $\fr=0$
 \cite{Dun05,Wil07,Gus11a}.
However, as shown in~\Figref{clustering_gravity}{\bf c} for $\st>\st^\ast$ caustics are less frequent.
For $\fr=0$ the rate of caustic formation is of the form ${\cal J}\sim {\rm e}^{-1/(3\ku^2\st)}$ in the white-noise limit provided that $\ku^2\st$ is small~\cite{Wil05}.
It follows from \Eqnref{KuEff} that caustics are deactivated when $\G$ becomes large, ${\rm e}^{-1/(C\ku_{\rm eff}^2\st)}$. 
{The constant $C$ can be determined by a WKB approximation, but in \Figref{clustering_gravity}{\bf c} its value was fitted.}
We note that caustics are rare when $\st^\ast$ is so small that the curve in Fig.~\ref{fig:clustering_gravity}{\bf a} turns before caustics are activated.

{\em Conclusions.}
We have derived a theory describing spatial clustering
of particles falling through a turbulent flow, making it possible to determine how the clustering depends on
the dimensionless parameters of the problem, $\fr$, $\st$, and $\ku$.
{Our theory clearly demonstrates that the inertial response
of the suspended particles to flow fluctuations and the effect of gravity
are not additive.}

For small and intermediate values of $\st$, our theory shows that particles
falling at finite $\fr$ cluster less, because correlations between particles and flow structures are destroyed.  
This explains earlier DNS results \cite{Pum04,Wan06,Fra07,Aya08a,Woi09}.  
% We find that 
% at very small values of $\st$, unless $\fr$ is very large, clustering due to the Maxey centrifuge effect is only slightly reduced,
% while clustering at intermediate values of $\st$ is reduced more.

For large values of $\st$, our calculations show that settling particles may cluster strongly. This is surprising
because for $F\!=\!0$, particles are uniformly distributed at large Stokes numbers. 
Our theory shows that when the particles fall rapidly enough,
they see the fluid gradients as a white-noise signal, giving
rise to substantial clustering by the mechanism of multiplicative amplification,
distinct from the mechanism causing clustering at small Stokes numbers (preferential concentration).
This surprising result is consistent with the observations made in \cite{Woi09}, but 
we note that in turbulent flows sweeping by large eddies may affect the dynamics of
rapidly falling particles, and the clustering of rapidly falling particles may be modified
by interactions with large eddies. 

We find that for large values of $\st$ the spatial clustering is strongly anisotropic because 
the separation vector between two neighbouring particles tends to align with $\pm\hat{ \ve g}$. 
Pad\'e{}-Borel resummation of high-order perturbation theory allows us to quantify this anisotropy.

Finally, we find that the rate of caustic formation is reduced when  $\st$
is large and $\fr$ not too small. This implies smaller relative velocities between identical close-by particles,
resulting in lower collision rates between such particles.

The problem of particles falling under gravity through a turbulent
aerosol is important for rain initiation in warm turbulent rain clouds \cite{Dev12}.
In this case the Stokes number takes values $\st\sim 10^{-3}(a/\mu{\rm m})^2$~\cite{Sha03}. The Stokes number increases as water droplets grow from $\st\sim 10^{-3}$ for small droplets (size $1\mu{\rm m}$)
to $\st\sim 10$ for large droplets ($100\,\mu{\rm m}$).
In the absence of gravity in a flow with $\ku\sim 1$, clustering is largest when $\st$ is of order unity.
Typical values of the characteristic scales in vigorously turbulent rain clouds are $\tau\sim 10\,{\rm ms}$, $\eta\sim 1\,{\rm mm}$ and $u_0\sim 0.1\,{\rm m}/{\rm s}$~\cite{Sha03} which gives $\ku$ and $\fr$ of order unity (blue curve in \Figref{clustering_gravity}).
Gravitational settling begins to become important for the motion of droplets larger than about $20\,\mu{\rm m}$~\cite{Jon96}. 
{The corresponding Stokes number, $\st\sim 0.4$,} is of the order of $\st^\ast$ [defined below Eq.~(\ref{eq:KuEff})] for vigorously turbulent rain clouds.
Our theory thus predicts that gravitational settling substantially changes the spatial clustering of rain droplets falling in turbulent rain clouds. 
This is expected to significantly increase the rate of  turbulence-induced collision-coalescence of droplets of similar sizes, while
the suppression of caustics at large $\st$ has the opposite effect.

{\em Acknowledgements}.
Financial support by Vetenskapsr\aa{}det and
by the G\"oran Gustafsson Foundation for Research in Natural Sciences and Medicine are gratefully acknowledged.
The numerical computations were performed using resources
provided by C3SE and SNIC.


\begin{thebibliography}{39}
\expandafter\ifx\csname natexlab\endcsname\relax\def\natexlab#1{#1}\fi
\expandafter\ifx\csname bibnamefont\endcsname\relax
  \def\bibnamefont#1{#1}\fi
\expandafter\ifx\csname bibfnamefont\endcsname\relax
  \def\bibfnamefont#1{#1}\fi
\expandafter\ifx\csname citenamefont\endcsname\relax
  \def\citenamefont#1{#1}\fi
\expandafter\ifx\csname url\endcsname\relax
  \def\url#1{\texttt{#1}}\fi
\expandafter\ifx\csname urlprefix\endcsname\relax\def\urlprefix{URL }\fi
\providecommand{\bibinfo}[2]{#2}
\providecommand{\eprint}[2][]{\url{#2}}

\bibitem[{\citenamefont{Saw et~al.}(2008)\citenamefont{Saw, Shaw,
  Ayyalasomayajula, Chuang, and Gylfason}}]{Saw08}
\bibinfo{author}{\bibfnamefont{E.-W.} \bibnamefont{Saw}},
  \bibinfo{author}{\bibfnamefont{R.}~\bibnamefont{Shaw}},
  \bibinfo{author}{\bibfnamefont{S.}~\bibnamefont{Ayyalasomayajula}},
  \bibinfo{author}{\bibfnamefont{P.}~\bibnamefont{Chuang}}, \bibnamefont{and}
  \bibinfo{author}{\bibfnamefont{A.}~\bibnamefont{Gylfason}},
  \bibinfo{journal}{Phys.~Rev.~Lett.} \textbf{\bibinfo{volume}{100}},
  \bibinfo{pages}{214501} (\bibinfo{year}{2008}).

\bibitem[{\citenamefont{Salazar et~al.}(2008)\citenamefont{Salazar, de~Jonag,
  Cao, Woodward, Meng, and Collins}}]{Sal08}
\bibinfo{author}{\bibfnamefont{J.}~\bibnamefont{Salazar}},
  \bibinfo{author}{\bibfnamefont{J.}~\bibnamefont{de~Jonag}},
  \bibinfo{author}{\bibfnamefont{L.}~\bibnamefont{Cao}},
  \bibinfo{author}{\bibfnamefont{S.~H.} \bibnamefont{Woodward}},
  \bibinfo{author}{\bibfnamefont{H.}~\bibnamefont{Meng}}, \bibnamefont{and}
  \bibinfo{author}{\bibfnamefont{L.}~\bibnamefont{Collins}},
  \bibinfo{journal}{J.~Fluid Mech.} \textbf{\bibinfo{volume}{600}},
  \bibinfo{pages}{245} (\bibinfo{year}{2008}).

\bibitem[{\citenamefont{Gibert et~al.}(2012)\citenamefont{Gibert, Xu, and
  Bodenschatz}}]{Gib12}
\bibinfo{author}{\bibfnamefont{M.}~\bibnamefont{Gibert}},
  \bibinfo{author}{\bibfnamefont{H.}~\bibnamefont{Xu}}, \bibnamefont{and}
  \bibinfo{author}{\bibfnamefont{E.}~\bibnamefont{Bodenschatz}},
  \bibinfo{journal}{J.~Fluid Mech.} \textbf{\bibinfo{volume}{698}},
  \bibinfo{pages}{160} (\bibinfo{year}{2012}).

\bibitem[{\citenamefont{Warhaft}(2009)}]{War09}
\bibinfo{author}{\bibfnamefont{Z.}~\bibnamefont{Warhaft}},
  \bibinfo{journal}{Fluid Dyn. Res.} \textbf{\bibinfo{volume}{41}},
  \bibinfo{pages}{011201} (\bibinfo{year}{2009}).

\bibitem[{\citenamefont{Chun et~al.}(2005)\citenamefont{Chun, Koch, Rani,
  Ahluwalia, and Collins}}]{Chu05}
\bibinfo{author}{\bibfnamefont{J.}~\bibnamefont{Chun}},
  \bibinfo{author}{\bibfnamefont{D.~L.} \bibnamefont{Koch}},
  \bibinfo{author}{\bibfnamefont{S.~L.} \bibnamefont{Rani}},
  \bibinfo{author}{\bibfnamefont{A.}~\bibnamefont{Ahluwalia}},
  \bibnamefont{and} \bibinfo{author}{\bibfnamefont{L.~R.}
  \bibnamefont{Collins}}, \bibinfo{journal}{J.~Fluid Mech.}
  \textbf{\bibinfo{volume}{536}}, \bibinfo{pages}{219} (\bibinfo{year}{2005}).

\bibitem[{\citenamefont{Bec et~al.}(2006)\citenamefont{Bec, Biferale, Boffetta,
  Cencini, Musacchio, and Toschi}}]{Bec06}
\bibinfo{author}{\bibfnamefont{J.}~\bibnamefont{Bec}},
  \bibinfo{author}{\bibfnamefont{L.}~\bibnamefont{Biferale}},
  \bibinfo{author}{\bibfnamefont{G.}~\bibnamefont{Boffetta}},
  \bibinfo{author}{\bibfnamefont{M.}~\bibnamefont{Cencini}},
  \bibinfo{author}{\bibfnamefont{S.}~\bibnamefont{Musacchio}},
  \bibnamefont{and} \bibinfo{author}{\bibfnamefont{F.}~\bibnamefont{Toschi}},
  \bibinfo{journal}{Phys.~Fluids} \textbf{\bibinfo{volume}{18}},
  \bibinfo{pages}{091702} (\bibinfo{year}{2006}).

\bibitem[{\citenamefont{Calzavarini et~al.}(2009)\citenamefont{Calzavarini,
  Cencini, Lohse, and Toschi}}]{Cal08}
\bibinfo{author}{\bibfnamefont{E.}~\bibnamefont{Calzavarini}},
  \bibinfo{author}{\bibfnamefont{M.}~\bibnamefont{Cencini}},
  \bibinfo{author}{\bibfnamefont{D.}~\bibnamefont{Lohse}}, \bibnamefont{and}
  \bibinfo{author}{\bibfnamefont{F.}~\bibnamefont{Toschi}},
  \bibinfo{journal}{Phys.~Rev.~Lett.} \textbf{\bibinfo{volume}{101}},
  \bibinfo{pages}{084504} (\bibinfo{year}{2009}).

\bibitem[{\citenamefont{Meneguz and Reeks}(2010)}]{Men10}
\bibinfo{author}{\bibfnamefont{E.}~\bibnamefont{Meneguz}} \bibnamefont{and}
  \bibinfo{author}{\bibfnamefont{M.}~\bibnamefont{Reeks}},
  \bibinfo{journal}{J.~Fluid Mech.} \textbf{\bibinfo{volume}{686}},
  \bibinfo{pages}{338} (\bibinfo{year}{2010}).

\bibitem[{\citenamefont{Bec}(2003)}]{Bec03b}
\bibinfo{author}{\bibfnamefont{J.}~\bibnamefont{Bec}},
  \bibinfo{journal}{Phys.~Fluids} \textbf{\bibinfo{volume}{15}},
  \bibinfo{pages}{81} (\bibinfo{year}{2003}).

\bibitem[{\citenamefont{Maxey}(1987)}]{Max87}
\bibinfo{author}{\bibfnamefont{M.~R.} \bibnamefont{Maxey}},
  \bibinfo{journal}{J.~Fluid Mech.} \textbf{\bibinfo{volume}{174}},
  \bibinfo{pages}{441} (\bibinfo{year}{1987}).

\bibitem[{\citenamefont{Duncan et~al.}(2005)\citenamefont{Duncan, Mehlig,
  \"Ostlund, and Wilkinson}}]{Dun05}
\bibinfo{author}{\bibfnamefont{K.}~\bibnamefont{Duncan}},
  \bibinfo{author}{\bibfnamefont{B.}~\bibnamefont{Mehlig}},
  \bibinfo{author}{\bibfnamefont{S.}~\bibnamefont{\"Ostlund}},
  \bibnamefont{and}
  \bibinfo{author}{\bibfnamefont{M.}~\bibnamefont{Wilkinson}},
  \bibinfo{journal}{Phys.~Rev.~Lett.} \textbf{\bibinfo{volume}{95}},
  \bibinfo{pages}{240602} (\bibinfo{year}{2005}).

\bibitem[{\citenamefont{Mehlig et~al.}(2005)\citenamefont{Mehlig, Wilkinson,
  Duncan, Weber, and Ljunggren}}]{Meh05}
\bibinfo{author}{\bibfnamefont{B.}~\bibnamefont{Mehlig}},
  \bibinfo{author}{\bibfnamefont{M.}~\bibnamefont{Wilkinson}},
  \bibinfo{author}{\bibfnamefont{K.}~\bibnamefont{Duncan}},
  \bibinfo{author}{\bibfnamefont{T.}~\bibnamefont{Weber}}, \bibnamefont{and}
  \bibinfo{author}{\bibfnamefont{M.}~\bibnamefont{Ljunggren}},
  \bibinfo{journal}{Phys.~Rev.~E} \textbf{\bibinfo{volume}{72}},
  \bibinfo{pages}{051104} (\bibinfo{year}{2005}).

\bibitem[{\citenamefont{Wilkinson et~al.}(2007)\citenamefont{Wilkinson, Mehlig,
  \"Ostlund, and Duncan}}]{Wil07}
\bibinfo{author}{\bibfnamefont{M.}~\bibnamefont{Wilkinson}},
  \bibinfo{author}{\bibfnamefont{B.}~\bibnamefont{Mehlig}},
  \bibinfo{author}{\bibfnamefont{S.}~\bibnamefont{\"Ostlund}},
  \bibnamefont{and} \bibinfo{author}{\bibfnamefont{K.~P.}
  \bibnamefont{Duncan}}, \bibinfo{journal}{Phys. Fluids}
  \textbf{\bibinfo{volume}{19}}, \bibinfo{pages}{113303(R)}
  (\bibinfo{year}{2007}).

\bibitem[{\citenamefont{Gustavsson and Mehlig}(2011{\natexlab{a}})}]{Gus11a}
\bibinfo{author}{\bibfnamefont{K.}~\bibnamefont{Gustavsson}} \bibnamefont{and}
  \bibinfo{author}{\bibfnamefont{B.}~\bibnamefont{Mehlig}},
  \bibinfo{journal}{Europhys.~Lett.} \textbf{\bibinfo{volume}{96}},
  \bibinfo{pages}{60012} (\bibinfo{year}{2011}{\natexlab{a}}).

\bibitem[{\citenamefont{Falkovich and Pumir}(2004)}]{Pum04}
\bibinfo{author}{\bibfnamefont{G.}~\bibnamefont{Falkovich}} \bibnamefont{and}
  \bibinfo{author}{\bibfnamefont{A.}~\bibnamefont{Pumir}},
  \bibinfo{journal}{Phys. Fluids} \textbf{\bibinfo{volume}{16}},
  \bibinfo{pages}{L47} (\bibinfo{year}{2004}).

\bibitem[{\citenamefont{Wang et~al.}(2006)\citenamefont{Wang, Ayala, Xue, and
  Grabowski}}]{Wan06}
\bibinfo{author}{\bibfnamefont{L.}~\bibnamefont{Wang}},
  \bibinfo{author}{\bibfnamefont{O.}~\bibnamefont{Ayala}},
  \bibinfo{author}{\bibfnamefont{Y.}~\bibnamefont{Xue}}, \bibnamefont{and}
  \bibinfo{author}{\bibfnamefont{W.}~\bibnamefont{Grabowski}},
  \bibinfo{journal}{J. Atmos. Sci.} \textbf{\bibinfo{volume}{63}},
  \bibinfo{pages}{2397} (\bibinfo{year}{2006}).

\bibitem[{\citenamefont{Franklin et~al.}(2007)\citenamefont{Franklin,
  Vaillancourt, and Yau}}]{Fra07}
\bibinfo{author}{\bibfnamefont{C.}~\bibnamefont{Franklin}},
  \bibinfo{author}{\bibfnamefont{P.}~\bibnamefont{Vaillancourt}},
  \bibnamefont{and} \bibinfo{author}{\bibfnamefont{M.}~\bibnamefont{Yau}},
  \bibinfo{journal}{J.~Atmos.~Sci.} \textbf{\bibinfo{volume}{64}},
  \bibinfo{pages}{938} (\bibinfo{year}{2007}).

\bibitem[{\citenamefont{Ayala et~al.}(2008)\citenamefont{Ayala, Rosa, Wang, and
  Grabowski}}]{Aya08a}
\bibinfo{author}{\bibfnamefont{O.}~\bibnamefont{Ayala}},
  \bibinfo{author}{\bibfnamefont{B.}~\bibnamefont{Rosa}},
  \bibinfo{author}{\bibfnamefont{L.-P.} \bibnamefont{Wang}}, \bibnamefont{and}
  \bibinfo{author}{\bibfnamefont{W.}~\bibnamefont{Grabowski}},
  \bibinfo{journal}{New~J.~Phys.} \textbf{\bibinfo{volume}{10}},
  \bibinfo{pages}{075015} (\bibinfo{year}{2008}).

\bibitem[{\citenamefont{Woittiez et~al.}(2009)\citenamefont{Woittiez, Jonker,
  and Portela}}]{Woi09}
\bibinfo{author}{\bibfnamefont{E.}~\bibnamefont{Woittiez}},
  \bibinfo{author}{\bibfnamefont{H.}~\bibnamefont{Jonker}}, \bibnamefont{and}
  \bibinfo{author}{\bibfnamefont{L.}~\bibnamefont{Portela}},
  \bibinfo{journal}{J. Atmos. Sci.} \textbf{\bibinfo{volume}{66}},
  \bibinfo{pages}{1926} (\bibinfo{year}{2009}).

\bibitem[{\citenamefont{Shaw}(2003)}]{Sha03}
\bibinfo{author}{\bibfnamefont{R.~A.} \bibnamefont{Shaw}},
  \bibinfo{journal}{Annu. Rev. Fluid Mech.} \textbf{\bibinfo{volume}{35}},
  \bibinfo{pages}{183} (\bibinfo{year}{2003}).

\bibitem[{\citenamefont{Wilkinson et~al.}(2005)\citenamefont{Wilkinson and Mehlig}}]{Wil05}
\bibinfo{author}{\bibfnamefont{M.}~\bibnamefont{Wilkinson}}, \bibnamefont{and}
  \bibinfo{author}{\bibfnamefont{B.}~\bibnamefont{Mehlig}},
  \bibinfo{journal}{Europhys.~Lett.} \textbf{\bibinfo{volume}{71}},
  \bibinfo{pages}{186} (\bibinfo{year}{2005}).

\bibitem[{\citenamefont{Wilkinson et~al.}(2006)\citenamefont{Wilkinson, Mehlig,
  and Bezuglyy}}]{Wil06}
\bibinfo{author}{\bibfnamefont{M.}~\bibnamefont{Wilkinson}},
  \bibinfo{author}{\bibfnamefont{B.}~\bibnamefont{Mehlig}}, \bibnamefont{and}
  \bibinfo{author}{\bibfnamefont{V.}~\bibnamefont{Bezuglyy}},
  \bibinfo{journal}{Phys.~Rev.~Lett.} \textbf{\bibinfo{volume}{97}},
  \bibinfo{pages}{048501} (\bibinfo{year}{2006}).

\bibitem[{\citenamefont{Falkovich et~al.}(2002)\citenamefont{Falkovich, Fouxon,
  and Stepanov}}]{Fal02}
\bibinfo{author}{\bibfnamefont{G.}~\bibnamefont{Falkovich}},
  \bibinfo{author}{\bibfnamefont{A.}~\bibnamefont{Fouxon}}, \bibnamefont{and}
  \bibinfo{author}{\bibfnamefont{G.}~\bibnamefont{Stepanov}},
  \bibinfo{journal}{Nature} \textbf{\bibinfo{volume}{419}},
  \bibinfo{pages}{151} (\bibinfo{year}{2002}).

\bibitem[{\citenamefont{Bec et~al.}(2010)\citenamefont{Bec, Biferale, Cencini,
  Lanotte, and Toschi}}]{Bec10}
\bibinfo{author}{\bibfnamefont{J.}~\bibnamefont{Bec}},
  \bibinfo{author}{\bibfnamefont{L.}~\bibnamefont{Biferale}},
  \bibinfo{author}{\bibfnamefont{M.}~\bibnamefont{Cencini}},
  \bibinfo{author}{\bibfnamefont{A.}~\bibnamefont{Lanotte}}, \bibnamefont{and}
  \bibinfo{author}{\bibfnamefont{F.}~\bibnamefont{Toschi}},
  \bibinfo{journal}{J.~Fluid Mech.} \textbf{\bibinfo{volume}{646}},
  \bibinfo{pages}{527} (\bibinfo{year}{2010}).

\bibitem[{\citenamefont{Gustavsson and Mehlig}(2011{\natexlab{b}})}]{Gus11b}
\bibinfo{author}{\bibfnamefont{K.}~\bibnamefont{Gustavsson}} \bibnamefont{and}
  \bibinfo{author}{\bibfnamefont{B.}~\bibnamefont{Mehlig}},
  \bibinfo{journal}{Phys.~Rev.~E} \textbf{\bibinfo{volume}{84}},
  \bibinfo{pages}{045304} (\bibinfo{year}{2011}{\natexlab{b}}).

\bibitem[{\citenamefont{Salazar and Collins}(2012)}]{Sal12}
\bibinfo{author}{\bibfnamefont{J.}~\bibnamefont{Salazar}} \bibnamefont{and}
  \bibinfo{author}{\bibfnamefont{L.}~\bibnamefont{Collins}},
  \bibinfo{journal}{J.~Fluid Mech.} \textbf{\bibinfo{volume}{696}},
  \bibinfo{pages}{45} (\bibinfo{year}{2012}).

\bibitem[{\citenamefont{Bewley et~al.}(2013)\citenamefont{Bewley, Saw, and
  Bodenschatz}}]{Bew13}
\bibinfo{author}{\bibfnamefont{G.~P.} \bibnamefont{Bewley}},
  \bibinfo{author}{\bibfnamefont{E.~W.} \bibnamefont{Saw}}, \bibnamefont{and}
  \bibinfo{author}{\bibfnamefont{E.}~\bibnamefont{Bodenschatz}},
  \bibinfo{journal}{New~J.~Phys.} \textbf{\bibinfo{volume}{15}},
  \bibinfo{pages}{083051} (\bibinfo{year}{2013}).

\bibitem[{\citenamefont{Gustavsson and Mehlig}(2014)}]{Gus14b}
\bibinfo{author}{\bibfnamefont{K.}~\bibnamefont{Gustavsson}} \bibnamefont{and}
  \bibinfo{author}{\bibfnamefont{B.}~\bibnamefont{Mehlig}},
  \bibinfo{journal}{Journal of Turbulence} {\textbf{\bibinfo{volume}{15}},
  \bibinfo{pages}{34}} (\bibinfo{year}{2014}).

\bibitem[{\citenamefont{Devenish et~al.}(2012)\citenamefont{Devenish, Bartello,
  Brenguier, Collins, Grabowski, IJzermans, Malinowski, Reeks, Vassilicos, Wang
  et~al.}}]{Dev12}
\bibinfo{author}{\bibfnamefont{B.~J.} \bibnamefont{Devenish}},
  \bibinfo{author}{\bibfnamefont{P.}~\bibnamefont{Bartello}},
  \bibinfo{author}{\bibfnamefont{J.-L.} \bibnamefont{Brenguier}},
  \bibinfo{author}{\bibfnamefont{L.~R.} \bibnamefont{Collins}},
  \bibinfo{author}{\bibfnamefont{W.~W.} \bibnamefont{Grabowski}},
  \bibinfo{author}{\bibfnamefont{R.~H.~A.} \bibnamefont{IJzermans}},
  \bibinfo{author}{\bibfnamefont{S.~P.} \bibnamefont{Malinowski}},
  \bibinfo{author}{\bibfnamefont{M.~W.} \bibnamefont{Reeks}},
  \bibinfo{author}{\bibfnamefont{J.~C.} \bibnamefont{Vassilicos}},
  \bibinfo{author}{\bibfnamefont{L.-P.} \bibnamefont{Wang}},
  \bibnamefont{et~al.}, \bibinfo{journal}{Q. J. R. Meteorol. Soc.}
  \textbf{\bibinfo{volume}{138}}, \bibinfo{pages}{1401} (\bibinfo{year}{2012}).

\bibitem[{\citenamefont{Gustavsson and Mehlig}(2013{\natexlab{b}})}]{Gus13a}
\bibinfo{author}{\bibfnamefont{K.}~\bibnamefont{Gustavsson}} \bibnamefont{and}
  \bibinfo{author}{\bibfnamefont{B.}~\bibnamefont{Mehlig}},
  \bibinfo{journal}{Phys.~Rev.~E} \textbf{\bibinfo{volume}{87}},
  \bibinfo{pages}{023016} (\bibinfo{year}{2013}{\natexlab{b}}).

\bibitem[{\citenamefont{Gustavsson et~al.}(2014)\citenamefont{Gustavsson,
  Einarsson, and Mehlig}}]{Gus13d}
\bibinfo{author}{\bibfnamefont{K.}~\bibnamefont{Gustavsson}},
  \bibinfo{author}{\bibfnamefont{J.}~\bibnamefont{Einarsson}},
  \bibnamefont{and} \bibinfo{author}{\bibfnamefont{B.}~\bibnamefont{Mehlig}},
  \bibinfo{journal}{Phys.~Rev.~Lett.} {\textbf{\bibinfo{volume}{112}},
  \bibinfo{pages}{014501}} (\bibinfo{year}{2014}).

\bibitem[{\citenamefont{Gustavsson et~al.}(2014)\citenamefont{Gustavsson,
  Vajedi, and Mehlig}}]{supp}
  \bibinfo{title}{See Supplemental Material at [URL will be inserted by publisher] for a summary of the method used to derive \Eqnref{lambdaSum_Gravity}.}).

\bibitem[{\citenamefont{Sommerer and Ott}(1993)}]{Som93}
\bibinfo{author}{\bibfnamefont{J.}~\bibnamefont{Sommerer}} \bibnamefont{and}
  \bibinfo{author}{\bibfnamefont{E.}~\bibnamefont{Ott}},
  \bibinfo{journal}{Science} \textbf{\bibinfo{volume}{259}},
  \bibinfo{pages}{334} (\bibinfo{year}{1993}).

\bibitem[{\citenamefont{Wilkinson and Mehlig}(2003)}]{Wil03}
\bibinfo{author}{\bibfnamefont{M.}~\bibnamefont{Wilkinson}} \bibnamefont{and}
  \bibinfo{author}{\bibfnamefont{B.}~\bibnamefont{Mehlig}},
  \bibinfo{journal}{Phys.~Rev.~E} \textbf{\bibinfo{volume}{68}},
  \bibinfo{pages}{040101(R)} (\bibinfo{year}{2003}).

\bibitem[{\citenamefont{Mehlig and Wilkinson}(2004)}]{Meh04}
\bibinfo{author}{\bibfnamefont{B.}~\bibnamefont{Mehlig}} \bibnamefont{and}
  \bibinfo{author}{\bibfnamefont{M.}~\bibnamefont{Wilkinson}},
  \bibinfo{journal}{Phys.~Rev.~Lett.} \textbf{\bibinfo{volume}{92}},
  \bibinfo{pages}{250602} (\bibinfo{year}{2004}).

\bibitem[{\citenamefont{Kaplan and Yorke}(1979)}]{Kap79}
\bibinfo{author}{\bibfnamefont{J.}~\bibnamefont{Kaplan}} \bibnamefont{and}
  \bibinfo{author}{\bibfnamefont{J.~A.} \bibnamefont{Yorke}},
  \bibinfo{journal}{Springer Lecture Notes in Mathematics}
  \textbf{\bibinfo{volume}{730}}, \bibinfo{pages}{204} (\bibinfo{year}{1979}).

\bibitem[{\citenamefont{Wilkinson et~al.}(2010)\citenamefont{Wilkinson, Mehlig,
  and Gustavsson}}]{Wil10b}
\bibinfo{author}{\bibfnamefont{M.}~\bibnamefont{Wilkinson}},
  \bibinfo{author}{\bibfnamefont{B.}~\bibnamefont{Mehlig}}, \bibnamefont{and}
  \bibinfo{author}{\bibfnamefont{K.}~\bibnamefont{Gustavsson}},
  \bibinfo{journal}{Europhys.~Lett.} \textbf{\bibinfo{volume}{89}},
  \bibinfo{pages}{50002} (\bibinfo{year}{2010}).

\bibitem[{\citenamefont{Bezuglyy et~al.}(2006)\citenamefont{Bezuglyy, Mehlig, Wilkinson, Nakamura, and Arvedson}}]{Bez06}
\bibinfo{author}{\bibfnamefont{V.}~\bibnamefont{Bezuglyy}},
\bibinfo{author}{\bibfnamefont{B.}~\bibnamefont{Mehlig}},
\bibinfo{author}{\bibfnamefont{M.}~\bibnamefont{Wilkinson}},
\bibinfo{author}{\bibfnamefont{K.}~\bibnamefont{Nakamura}}, \bibnamefont{and}
\bibinfo{author}{\bibfnamefont{E.}~\bibnamefont{Arvedson}}, \bibinfo{journal}{J. Math. Phys.} \textbf{\bibinfo{volume}{47}},
  \bibinfo{pages}{073301} (\bibinfo{year}{2006}).

\bibitem[{\citenamefont{Fouxon et~al.}(2008)\citenamefont{Fouxon and Horvai}}]{Fou08}
\bibinfo{author}{\bibfnamefont{I.}~\bibnamefont{Fouxon}},  \bibnamefont{and}
  \bibinfo{author}{\bibfnamefont{P.}~\bibnamefont{Horvai}}, \bibinfo{journal}{Phys. Rev. Lett.} \textbf{\bibinfo{volume}{100}},
  \bibinfo{pages}{040601} (\bibinfo{year}{2008}).

\bibitem[{\citenamefont{Jonas}(1996)}]{Jon96}
\bibinfo{author}{\bibfnamefont{P.~R.} \bibnamefont{Jonas}},
  \bibinfo{journal}{Atmos. Res.} \textbf{\bibinfo{volume}{40}},
  \bibinfo{pages}{283} (\bibinfo{year}{1996}).

\end{thebibliography}
\end{document}

\title{Supplemental material to \lq Clustering of particles falling in a turbulent flow\rq}
\author{K. Gustavsson, S. Vajedi, and B. Mehlig}
\affiliation{Department of Physics, Gothenburg University, 41296 Gothenburg, Sweden}

\maketitle

\section{Perturbation series for the preferential concentration}
In the  Letter the average divergence of the particle velocity field in the steady state, $\langle\ve \nabla \cdot \ve v\rangle_\infty$, is derived using the 
perturbative method described in Refs.~[14,29,30] in the Letter.
This calculation is outlined in the Letter, but for convenience we provide more substantial information in this supplemental document.

We model the motion of an individual particle using Eq. (1) in the Letter
\begin{align}
\eqnlab{eqmStokesGravity_r}
\dot{\ve r}&=\ku\ve v\\
\dot{\ve v}&=(\ve u(\ve r,t)-\ve v)/\st+\fr\hat{\ve g}\,.
\eqnlab{eqmStokesGravity_v}
\end{align}
We want to evaluate $\ve \nabla \cdot \ve v$ taking into account of the local environment experienced by a particle following a trajectory $\ve r_t$. Here $\ve r_t$ is a solution to Eqs.~\eqnref{eqmStokesGravity_r} and~\eqnref{eqmStokesGravity_v}.
In order to do this we consider the dynamics of the particle-velocity gradient matrix $\ma Z$ with components $Z_{ij}=\partial v_i/\partial r_j$ evaluated along the particle trajectory $\ve r_t$. We note that $\ve \nabla \cdot \ve v=\tr\ma Z$.
The time evolution of $\ma Z$ is $\dot{\ma Z}={\rm d}[\ve\nabla\ve v\T]/{\rm d}t=\ve\nabla\dot{\ve v}\T-\ma Z\ve\nabla\dot{\ve r}\T$, which is determined by Eqs.~\eqnref{eqmStokesGravity_r} and~\eqnref{eqmStokesGravity_v}
\begin{align}
\dot{\ma Z}&=(\ma A(\ve r,t)-\ma Z)/\st-\ku\ma Z^2\,.
\eqnlab{eqmZ}
\end{align}
Here $\ma A$ is the fluid-velocity gradient matrix with components $A_{ij}=\partial u_i/\partial r_j$.
To solve \Eqnref{eqmZ} we assume that the Kubo number $\ku$ is small. This allows for a series expansion of $\ma Z$ and $\ma A$ in terms of $\ku$ in \Eqnref{eqmZ}.
An implicit solution to \Eqnref{eqmZ} is given by
\begin{align}
\ma Z(\ve r_t,t)&=\ma Z_0e^{-t/\st}+\int_0^t{\rm d}t_1e^{(t_1-t)/\st}\left[\frac{1}{\st}\ma A(\ve r_{t_1},t_1)-\ku\ma Z(\ve r_{t_1},t_1)^2\right]\,.
\end{align}
By iteratively substituting this expression for $\ma Z$ into the right hand side we recursively determine $\ma Z$ in terms of its initial value $\ma Z_0$ and of $\ma A$ to any desired order in $\ku$
\begin{align}
\ma Z(\ve r_t,t)&=\ma Z_0e^{-t/\st}
+\frac{1}{\st}\int_0^t{\rm d}t_1e^{(t_1-t)/\st}\ma A(\ve r_{t_1},t_1)
\nn\\&
-\ku\int_0^t{\rm d}t_1e^{(t_1-t)/\st}\left[\ma Z_0e^{-t_1/\st}+\frac{1}{\st}\int_0^{t_1}{\rm d}t_2e^{(t_2-t_1)/\st}\ma A(\ve r_{t_2},t_2)\right]^2
+\dots\,.
\eqnlab{ZsolA}
\end{align}
This solution depends implicitly on the gravity parameter $\fr$  because the $\ma A$-matrices are evaluated at the actual particle paths,
affected by gravitational settling.
In order to expand $\ma A(\ve r_t,t)$ in terms of small values of $\ku$, we first make an expansion  of \Eqnref{eqmStokesGravity_r} in terms of small fluctuations in the velocity field $\ve u$ (small values of $\ku$) around deterministic trajectories $\ve r_t^{(\rm d)}$.
Eqs.~\eqnref{eqmStokesGravity_r} and~\eqnref{eqmStokesGravity_v} cannot be solved explicitly because of the dependence on the particle trajectory $\ve r_t$ in the fluid velocity field $\ve u(\ve r_t,t)$.
\Eqnref{eqmStokesGravity_r} has an implicit solution
\begin{align}
\ve r_t&=\ve r_t^{(\rm d)}+\frac{\ku}{\st}\int_0^t{\rm d}t_1\int_0^{t_1}{\rm d}t_2e^{-(t_1-t_2)/\st}\ve u(\ve r_{t_2},t_2)\,,
\eqnlab{rsol_implicit}
\end{align}
with deterministic part $\ve r_t^{(\rm d)}$ given by Eq.~(2) in the Letter,
\begin{align}
\ve r_t^{(\rm d)}=\ve r_0+\ku\ve v_{\rm s} t+\ku\st(\ve v_0-\ve v_{\rm s})(1-e^{-t/\st})\,.
\eqnlab{rdet}
\end{align}
Here $\ve v_s\equiv \fr\st\hat{\ve g}$ is the settling velocity of \Eqnref{eqmStokesGravity_v} with $\ve u=0$.

We assume that the deviation $\delta\ve r_t\equiv \ve r_t-\ve r_t^{(\rm d)}$ from the deterministic solution $\ve r_t^{(\rm d)}$ in \Eqnref{rsol_implicit} is small. This assumption is always true for short enough times, but it is also true for times much larger than the correlation time of the flow if the Kubo number is small enough as we assume here.
We make a Taylor expansion of $\ve u$ around the deterministic trajectories
\begin{align}
\ve u(\ve r_t,t)=\ve u(\ve r_t^{(\rm d)},t)+[\delta\ve r_t\cdot\ve\nabla]\ve u(\ve r_t^{(\rm d)},t)+\dots\,.
\eqnlab{u_expansion}
\end{align}
Iteratively substituting $\delta\ve r_t=\ve r_t-\ve r_t^{(\rm d)}$ from \Eqnref{rsol_implicit} and $\ve u$ from \Eqnref{u_expansion} into \Eqnref{u_expansion} we find an expansion of $\ve u(\ve r_t,t)$ in terms of $\ve u(\ve r_t^{(\rm d)},t)$ and derivatives thereof. This expansion we cut off at some desired order in $\ku$.
Finally, by expanding $\ma A(\ve r_t,t)$ around $\ve r_t^{(\rm d)}$ as in \Eqnref{u_expansion} and inserting this expansion into the expansion for $\ma Z$, \Eqnref{ZsolA}, we obtain
an expansion of $\tr\ma Z$ in terms of a sum of products of $\ve u(\ve r_t^{(\rm d)},t)$ and derivatives thereof evaluated at the deterministic trajectory at different times.

To find the steady-state average $\langle\tr\ma Z\rangle_\infty$, we average all terms using the known statistics of the smooth model flow.
We expect the steady-state statistics of $\ma Z$ to be independent of the initial condition $\ve v_0$, $\ma Z_0$ and the initial flow configuration. We therefore let $\ve v_0=\ve v_{\rm s}$ and $\ma Z_0=0$. With this initial condition the deterministic trajectory \Eqnref{rdet} simplifies to $\ve r_t^{(\rm d)}=\G t$.

In the Letter we use a Gaussian-distributed random velocity field $\ve u$ in two spatial dimensions.
Write $\ve u=\ve\nabla\phi\wedge\hat{\ve e}_3/\sqrt{2}$, where $\hat{\ve e}_3$ is a unit vector in the $z$-direction.
We take the stream function $\phi$ to have zero mean $\langle \phi(\ve r,t)\rangle=0$ and correlation function
\begin{align}
\langle \phi(\ve r_1,t_1)\phi(\ve r_2,t_2)\rangle
\sim e^{-|t_1-t_2|-(\sve r_1-\sve r_2)^2/2}\,.
\eqnlab{corrfun_phi_def}
\end{align}
Inserting this correlation function into the  expansion for $\tr\ma Z$, we express the average $\langle\tr\ma Z\rangle_\infty$ in terms of multi-dimensional integrals over spatial derivatives of the correlation function, \Eqnref{corrfun_phi_def}.
Repeated partial integration allows us to express the expansion in terms of two-dimensional integrals.
We work out the large-time asymptotic behavior of these integrals which allows us to find the average divergence of the particle velocity field $\langle\ve \nabla \cdot \ve v\rangle_\infty=\langle\tr\ma Z\rangle_\infty$ (Eq.~(4) in the Letter)
\begin{align}
&\langle\ve \nabla \cdot \ve v\rangle_\infty=\frac{3\ku^3}{4\st^5\G^8}\bigg\{
-\sqrt{2}\G\st^2(13+17\st+15\st^2+3\st^3+\G^2\st^2(4-\st-3\st^2)+\G^4\st^4)\E\Big[\frac{1+\st}{\sqrt{2}\st\G}\Big]
\nn\\&
+2\G^2\st^3(5+4\st+3\st^2-\G^2\st^2(1+\st))
+(1+\st)^3(2(1+\st)^2-\G^2\st^2(\st-3))\E\left[\frac{1+\st}{\sqrt{2}\st\G}\right]^2
\nn\\&
-4\G\st(1+\st^2(2+\st^2+\G^2))\E\Big[\frac{1}{\G}\Big]
-2\sqrt{\pi}(1+\st^2)\G(-2+\st^2(-2+(-3+\st^2)\G^2))\I[\G,\st]\,.
\eqnlab{lambdaSum_Gravity}
\end{align}
Here
\begin{align}
\eqnlab{Edef}
\I[\G,\st]&\equiv\int_0^\infty{\rm d}t\exp\left[\frac{1}{\G^2}-\frac{t}{\st}-\frac{\G^2t^2}{4}\right]\erfc\left[\frac{1}{\G}+\frac{\G t}{2}\right]
\bigg\}\,,\\
\E[x]&\equiv \sqrt{\pi}e^{x^2}\erfc(x)\,,
\eqnlab{Idef}
\end{align}
and $\G\equiv\ku\st
\fr$ is an independent parameter ($\ku$ is small but $\G$ can take any value).

In the following we show how to evaluate the limiting behaviours of \Eqnref{lambdaSum_Gravity} for small/large values of $\st$ and $\G$ displayed in the Letter.
To expand \Eqnref{lambdaSum_Gravity} for small values of $\st$ or $\G$ we note that the asymptotic behaviour of $\E[x]$ in \Eqnref{Edef} for large values of $x$ is
\begin{align}
\E[x\gg 1]=\sum_{i=1}^{\infty} (2i- 3)!!(-2)^{1-i}x^{1-2i}=
\frac{1}{x} - \frac{1}{2x^3} + \frac{3}{4x^5} -\frac{15}{8x^7} + \frac{105}{16x^9}+ \dots\,.
\eqnlab{Elargex}
\end{align}
To expand the integral $\I[\G,\st]$ in \Eqnref{Idef} for small values of $\st$ we first rescale $t\to t\st$. Then we expand the resulting integrand in $\st$ and finally we integrate the expanded integrand.
We find
\begin{align}
\I[\G,\st\ll 1]&=
\st\int_0^\infty{\rm d}t\exp\left[\frac{1}{\G^2}-t-\frac{\G^2\st^2t^2}{4}\right]\erfc\left[\frac{1}{\G}+\frac{\G\st t}{2}\right]
\nn\\&
=\frac{\st}{\sqrt{\pi}}\int_0^\infty{\rm d}te^{-t}\bigg\{
\E\left[\frac{1}{\G}\right] \left(1-\frac{\G^2\st^2t^2}{4} + \frac{\G^4\st^4t^4}{32} - \frac{\G^6\st^6 t^6}{384}\right)
\nn\\&
\hspace{3cm}+\G\st t \bigg(-1 + \frac{\st t}{2} + (2\G^2-1)\frac{\st^2t^2}{6} - (9\G^2-2)\frac{\st^3 t^3}{48}
\nn\\&
\hspace{4.5cm} - (1 - 8\G^2 + 7\G^4)\frac{\st^4 t^4}{120} + \left(1- \frac{25\G^2}{2} + \frac{105\G^4}{4}\right)\frac{\st^5 t^5}{2880}\bigg)
\bigg\}+\dots
\nn\\&
=
\frac{\st}{\sqrt{\pi}}\bigg\{
\E\left[\frac{1}{\G}\right] \left(1 - \frac{\G^2\st^2}{2} + \frac{3\G^4\st^4}{4} - \frac{15\G^6\st^6}{8}\right)
+\G\st \bigg(-1 + \st + (2\G^2-1)\st^2
\nn\\&
\hspace{2cm}
+ \left(1 -\frac{ 9\G^2}{2}\right)\st^3 - (1 - 8\G^2 + 7\G^4)\st^4 + (4 - 50\G^2 + 105\G^4)\st^5\bigg)
\bigg\}+\dots .
\eqnlab{IsmallSt}
\end{align}
Inserting \Eqnref{IsmallSt} and \Eqnref{Elargex} to order $x^{-5}$ with $x=(1+\st)/(\sqrt{2}\st\G)$ into \Eqnref{lambdaSum_Gravity} and throwing away all terms of order higher than $\st^2$ gives the small-$\st$ limit of \Eqnref{lambdaSum_Gravity} quoted in Eq.~(5) in the Letter:
\begin{align}
\langle\ve \nabla \cdot \ve v\rangle_\infty &
\sim\frac{3\ku^3\st^2}{4\G^5}
(4\G-6\G^3-(4-4\G^2+3\G^4)\E[\G^{-1}])\,.
\eqnlab{lambdaSum_smallSt}
\end{align}

For small values of $\G$ we expand the integrand in \Eqnref{Idef} before integrating to obtain
\begin{align}
\I[\G\ll1,\st]&=
\int_0^\infty{\rm d}t\exp\left[\frac{1}{\G^2}-t/\st-\frac{\G^2t^2}{4}\right]\erfc\left[\frac{1}{\G}+\frac{\G t}{2}\right]
\nn\\&
=\int_0^\infty{\rm d}t
e^{-(1+\st)t/\st}\frac{\G}{\sqrt{\pi}}\bigg\{
1 -  (1 + t + t^2)\frac{\G^2}{2} + (6 + 6 t + 4 t^2 + 2 t^3 + t^4)\frac{\G^4}{8}
\nn\\&
\hspace{3.5cm}- (90 + 90 t + 54 t^2 + 24 t^3 + 9 t^4 + 3 t^5 + t^6)\frac{\G^6}{48}
\bigg\}+\dots
\nn\\&
=\frac{\G\st}{\sqrt{\pi}(1+\st)}\bigg\{
1 - \frac{\G^2 (1 + 3 \st + 4 \st^2)}{2(1 + \st)^2}
+ \frac{\G^4(3 + 15 \st + 31 \st^2 + 35 \st^3 + 28 \st^4)}{4(1+\st)^4}
\nn\\&
\hspace{2.5cm}
 - \frac{3\G^6(1+\st^2) (5 + 35 \st + 101 \st^2 + 147 \st^3 + 96 \st^4)}{8(1+\st)^6}
\nn\\&
\hspace{-1.5cm}
+\frac{3\G^8(35 + 315 \st + 1265 \st^2 + 2985 \st^3 + 4593 \st^4 + 4857 \st^5 +  3683 \st^6 + 2187 \st^7 + 1328 \st^8)}{16(1 +  \st)^8}
\bigg\}+\dots\,.
\eqnlab{IsmallG}
\end{align}
Inserting \Eqnref{IsmallG} and \Eqnref{Elargex} with $x=(1+\st)/(\sqrt{2}\st\G)$ and $x=1/\G$ into \Eqnref{lambdaSum_Gravity} and discarding all terms of order higher than $\G^2$ gives Eq.~(6) in the Letter:
\begin{align}
&\langle\ve \nabla \cdot \ve v\rangle_\infty
 = -6\ku^3\st^2\frac{1+3\st+\st^2}{(1+\st)^3}
+9\ku^3\G^2\st^2\frac{1+5\st+12\st^2+20\st^3+4\st^4}{(1 + \st)^5}
+\dots\,.
\end{align}

In the limit of both $\st$ and $\G$ being large, we use the lowest-order approximation of $\E[x]$ for small values of $x$
\begin{align}
\E[x\ll 1]\sim\sqrt{\pi}\,.
\end{align}
When $\st,\G\gg 1$ we also have
\begin{align}
\eqnlab{Edef1}
\I[\G\gg 1,\st\gg 1]&\sim\int_0^\infty{\rm d}t\exp\left[-\frac{\G^2t^2}{4}\right]\erfc\left[\frac{\G t}{2}\right]=\frac{\sqrt{\pi}}{2\G}\,.
\end{align}
Inserting these relations into \Eqnref{lambdaSum_Gravity} we find that the first term dominates with the asymptotic behavior of Eq.~(7) in the Letter:
\begin{align}
&\langle\ve \nabla \cdot \ve v\rangle_\infty\sim-\sqrt{2\pi}\frac{3\ku^3\st}{4\G^3}\,.
\end{align}

\section{Perturbation series for the alignment of particle pairs with gravity}
We compute the alignment of $\hat{\ve R}$ with $\hat{\ve g}$ using the method described in the previous section.
Linearisation of \Eqnref{eqmStokesGravity_r} gives $\dot{\ve R}=\ku\ma Z\ve R$ for infinitesimally small separations $\ve R$.
From this linearisation it follows that
\begin{align}
\frac{{\rm d}\hat{\ve R}}{{\rm d}t}=\ku[\ma Z\hat{\ve R}-(\hat{\ve R}{}\T\ma Z\hat{\ve R})\hat{\ve R}]\,.
\end{align}
This equation we solve implicitly for $\hat{\ve R}$
\begin{align}
\hat{\ve R}(\ve r_t,t)=\hat{\ve R}_0+\ku\int_0^t{\rm d}t_1[\ma Z(\ve r_{t_1},t_1)\hat{\ve R}(\ve r_{t_1},t_1)-(\hat{\ve R}{}\T(\ve r_{t_1},t_1)\ma Z(\ve r_{t_1},t_1)\hat{\ve R}(\ve r_{t_1},t_1))\hat{\ve R}(\ve r_{t_1},t_1)]\,.
\eqnlab{Rimplicit}
\end{align}
We expand this solution in terms of small values of $\ku$ by inserting the expansion for $\ma Z$ obtained in the previous section, and by repeatedly inserting the solution $\hat{\ve R}(\ve r_t,t)$ into the right hand side of \Eqnref{Rimplicit}.
This gives $\hat{\ve R}(\ve r_t,t)$ in terms of the initial direction $\hat{\ve R}_0$ and of spatial derivatives of the flow evaluated at $\ve r_r^{(\rm d)}$.
Multiplying this expansion with $\hat{\ve g}$ and raising the resulting expression to even powers, we obtain an expansion for $(\hat{\ve R}\cdot\hat{\ve g})^{2p}$ with $p=0,1,\dots$.
After taking an average, we find an expression for the even moments $\langle(\hat{\ve R}\cdot\hat{\ve g})^{2p}\rangle_\infty$
which involves secular terms and terms which depend on the initial configuration through $(\hat{\ve R}_0\cdot\hat{\ve g})^{2p}$, $\ve v_0$, $\ma Z_0$, $\ve u_0$, etc.
To lowest non-trivial order in $\ku$ the secular term takes the form
\begin{align}
\langle&(\hat{\ve R}\cdot\gghat)^{2p}\rangle_{\rm secular}
=t\frac{\ku^2p}{2\G^5}(\nog)^{2p-2}\bigg\{
p\Big[4(\nog)^2((\nog)^2-1)\G((\nog)^2+\G^2(5(\nog)^2-6))
\nn\\&
- 2((\nog)^2-1)((\nog)^4+6\G^2(\nog)^2((\nog)^2-1)+3\G^4((\nog)^2-1)^2) \sqrt{2}\E\left[\frac{1}{\sqrt{2}\G}\right]
\Big]
\nn\\&
+ 2(\nog)^4\G(-1+2(\nog)^2+\G^2(-11+10(\nog)^2))
\nn\\&
- (12\G^2(\nog)^4 ((\nog)^2-1) + (\nog)^4(2(\nog)^2-1) + \G^4(3 - 9(\nog)^4 + 6(\nog)^6)) \sqrt{2}\E\left[\frac{1}{\sqrt{2}\G}\right]
\bigg\}
\eqnlab{ngSecular}
\end{align}
Now we average over $\nog$ and require that $\langle(\hat{\ve R}\cdot\gghat)^{2p}\rangle_{\rm secular}$ vanishes to ensure that the steady state is correctly described (this condition also ensures that the lowest-order dependence on the initial configuration $\ma Z_0$, $\ve u_0$, $\ma A_0$, \dots vanishes).
This procedure results in a recursive relation for $\langle(\hat{\ve R}_0\cdot\hat{\ve g})^{2p}\rangle_\infty$ in $p$. 
We solve this recursion using an expansion in  $\G$, incorporating the boundary condition that 
\begin{align}
\langle(\hat{\ve R}_0\cdot\hat{\ve g})^0\rangle_\infty=1\,.
\eqnlab{BC}
\end{align}
We expand $\langle(\nog)^{2p}\rangle_\infty=\sum_{i=0}^\infty c^{(2i)}_{2p}\G^{2i}$ [terms that are odd in $\G$ turn out to vanish due to the 
boundary condition  \eqnref{BC}].
This expansion inserted in \Eqnref{ngSecular} gives solvable recursion relations for $c^{(2i)}_{2p}$ for each value of $i$. The lowest order equations ($i=0$ and $i=1$, the equations for negative values of $i$ are automatically satisfied due to the form of the expansion of $\E[1/(\sqrt{2}\G)]$) are
\begin{align}
0&=2pc^{(0)}_{2p}+(1-2p)c^{(0)}_{2p-2}\,,\\
0&=2pc^{(2)}_{2p}+(1-2p)c^{(2)}_{2p-2}-4(1+2p)c^{(0)}_{2p+2}+6pc^{(0)}_{2p}+(2p-1)c^{(0)}_{2p-2}\,.
\eqnlab{recursion}
\end{align}
The first equation has the solution $c^{(0)}_{2p}=A_{0}(2p-1)!!/(2^pp!)$, where the recursion coefficient $A_{0}=1$ is determined by the boundary condition \eqnref{BC}, i.e. $c^{(2i)}_{0}=\delta_{i,0}$. This gives $c^{(0)}_{2p}=(2p-1)!!/(2^pp!)$, which is identical to the result obtained by isotropically distributed directions $\hat{\ve R}_0$.
Insert this solution into \Eqnref{recursion} and solve the resulting recursion to find
$c^{(2)}_{2p}=((A_1 + p(1+A_1))(2p-1)!!)/( 2^{p}(p+1)!)$. With $A_1=0$ from the boundary condition \eqnref{BC}, we find the second term in Eq.~(8) in the  Letter.
By repeating this procedure to calculate $c^{(2i)}_{p}$ for larger values of $i$ we obtain an asymptotically divergent series (the first few coefficients are tabulated in Table~1).
We use Pad\'e-Borel resummation to evaluate this series. For the $40$ first coefficients this summation works well up to $\G\sim 10$. Iin Fig. 2b in the Letter the resummed series up to $\G=3$ is shown.
\begin{table}
\begin{tabular}{|c|llll|}
\hline
\backslashbox{$i$}{$p$} & 1 & 2 & 3 & 4 \cr
\hline
0 & 1/2 & 3/8 & 5/16 & 35/128 \cr
1 & 1/4 & 1/4 & 15/64 & 7/32 \cr
2 & -5/4 & -79/64 & -147/128 & -273/256 \cr
3 & 339/32 & 21/2 & 1253/128 & 1165/128 \cr
4 & -3825/32 & -15269/128 & -28581/256 & -1704941/16384 \cr
5 & 104879/64 & 105409/64 & 12678519/8192 & 2961973/2048 \cr
6 & -1668951/64 & -216094431/8192 & -407618699/16384 & -763641419/32768 \cr
7 & 1920979179/4096 & 977072365/2048 & 7396072887/16384 & 6942522817/16384 \cr
8 & -38379090493/4096 & -156920472515/16384 & -297784337115/32768 & -4480484018323/524288 \cr
9 & 1683355599405/8192 & 1727719300533/8192 & 52582881150441/262144 & 24762276181371/131072 \cr
10 & -40167158309193/8192\; & -1323779050232049/262144\; & -2523129742965693/524288\; & -4759032084160995/1048576\; \cr
\hline
\end{tabular}
\caption{The eleven first non-zero coefficients of $c^{(2i)}_{2p}$ used in Pad\'e-Borel summation to plot the solid lines in Fig. 2{\bf b} in the  Letter.}
\end{table}